\documentclass[journal]{vgtc}                
\ifpdf
  \pdfoutput=1\relax                   
  \usepackage{graphicx}                
  \DeclareGraphicsExtensions{.pdf,.png,.jpg,.jpeg} 
\else
  \usepackage{graphicx}                
  \DeclareGraphicsExtensions{.eps}     
\fi%

\graphicspath{{figures/}{pictures/}{images/}{./}} 

\usepackage{microtype}                 
\PassOptionsToPackage{warn}{textcomp}  
\usepackage{textcomp}                  
\usepackage{mathptmx}                  
\usepackage{times}                     
\usepackage{cite}                      
\usepackage{tabu}                      
\usepackage{booktabs}                  

\usepackage{enumitem}
\setlist[itemize]{noitemsep, topsep=0pt}
\usepackage{wrapfig}
\usepackage{caption}
\usepackage{titlecaps}
\usepackage{dblfloatfix}
\usepackage{amsmath}
\usepackage{amssymb}
\usepackage{newtxmath}
\usepackage{fontawesome}

\newcommand{\cmo}{\textcolor[rgb]{0, 0, 0}}
\newcommand{\rom}[1]{\uppercase\expandafter{\romannumeral #1\relax}}
\newcommand{\para}[1]{\vspace{1mm}\noindent\textbf{#1}}
\newcommand{\quot}[1]{\emph{``#1''}}
\newcommand{\numClip}{155}
\newcommand{\numVideo}{30}
\newcommand{\numMinutes}{3600}
\newcommand{\system}{Sporthesia}

\newcommand{\eg}{\emph{e.g.}}
\newcommand{\ie}{\emph{i.e.}}

\newcommand{\obj}{\texttt{object}}
\newcommand{\objs}{\texttt{objects}}
\newcommand{\act}{\texttt{action}}
\newcommand{\acts}{\texttt{actions}}
\newcommand{\data}{\texttt{data}}
\newcommand{\emo}{\texttt{emotional cue}}
\newcommand{\emos}{\texttt{emotional cues}}

\newcommand{\Obj}{\texttt{Object}}
\newcommand{\Objs}{\texttt{Objects}}
\newcommand{\Act}{\texttt{Action}}
\newcommand{\Acts}{\texttt{Actions}}
\newcommand{\Data}{\texttt{Data}}

\newcommand{\Emos}{\texttt{Emotional Cues}}

\onlineid{1238}

\vgtccategory{Research}
\vgtcpapertype{Systems/Technique}

\title{\system{}: Augmenting Sports Videos Using Natural Language}

\author{Chen Zhu-Tian, Qisen Yang, Xiao Xie, Johanna Beyer, Haijun Xia, Yingcai Wu, and Hanspeter Pfister}
\authorfooter{
\item Chen Zhu-Tian, Johanna Beyer, and Hanspeter Pfister are with with John
A. Paulson School of Engineering and Applied Sciences, Harvard University,
Cambridge, MA. E-mail: \{ztchen, jbeyer, pfister\}@g.harvard.edu.

    \item Qisen Yang, Xiao Xie, and Yingcai Wu are with Zhejiang University. E-mail: \{qs\_yang, xxie, ycwu\}@zju.edu.cn. This work was done when Qisen Yang was an intern at Harvard University.

    \item Haijun Xia is with UC San Diego. E-mail: haijunxia@ucsd.edu.
}

\abstract{Augmented sports videos, which combine visualizations and video effects to present data in actual scenes, can communicate insights engagingly and thus have been increasingly popular for sports enthusiasts around the world. 
Yet, creating augmented sports videos remains a challenging task, requiring considerable time and video editing skills. 
On the other hand, sports insights are often communicated using natural language, such as in commentaries, oral presentations, and articles,
but usually lack visual cues.
Thus, this work aims to facilitate the creation of augmented sports videos by enabling analysts to directly create visualizations embedded in videos using insights expressed in natural language. To achieve this goal, we propose a three-step approach – 1) detecting visualizable entities in the text, 2) mapping these entities into visualizations, and 3) scheduling these visualizations to play with the video – and analyzed 155 sports video clips and the accompanying commentaries for accomplishing these steps. 
Informed by our analysis, we have designed and implemented Sporthesia, a proof-of-concept system that 
takes racket-based sports videos and textual commentaries as the input and outputs augmented videos.
We demonstrate Sporthesia’s applicability in two exemplar scenarios, \ie, authoring augmented sports videos using text and augmenting historical sports videos based on auditory comments. 
A technical evaluation shows that Sporthesia achieves high accuracy (F1-score of 0.9) in detecting visualizable entities in the text. 
An expert evaluation with eight sports analysts suggests high utility, effectiveness, and satisfaction with our language-driven authoring method and provides insights for future improvement and opportunities.
}


\CCScatlist{ 
 \CCScat{K.6.1}{Management of Computing and Information Systems}%
{Project and People Management}{Life Cycle};
 \CCScat{K.7.m}{The Computing Profession}{Miscellaneous}{Ethics}
}

\teaser{
  \centering
  \hspace*{-1.2cm}
  \includegraphics[width=1.15\linewidth]{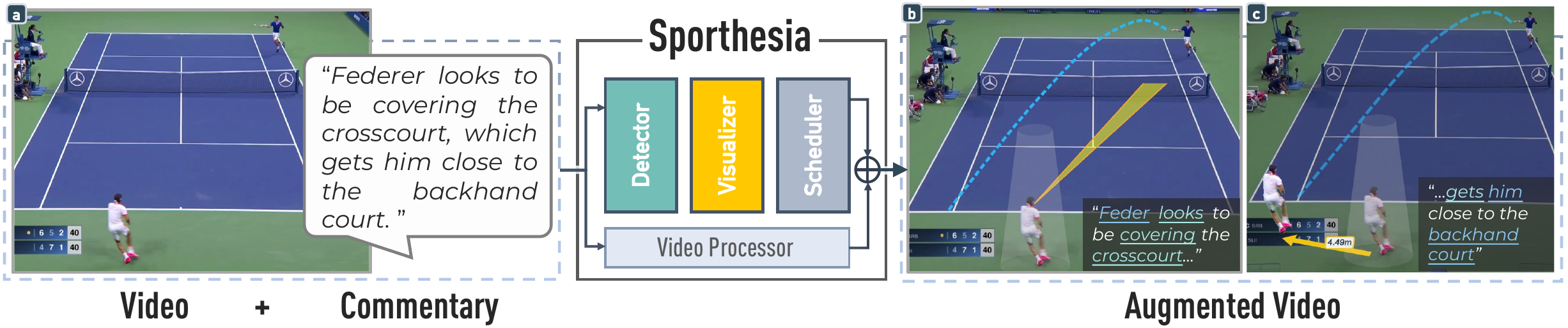}
  \caption{\system{} takes raw video footage and commentary text of racket-based sports as input, and outputs an augmented video.
  To achieve this,
  three key steps are taken: 1) detecting the visualizable entities in the text,
  2) mapping the entities to visualizations, and 3) scheduling the visualizations to play with the raw video.
  }
  \label{fig:teaser}
}

\renewcommand{\manuscriptnotetxt}{}

\vgtcinsertpkg

\begin{document}


\firstsection{Introduction}

\firstsection{Introduction}

\maketitle

Augmented sports videos are becoming increasingly popular as a form to present sports data. 
In recent years, 
an increasing amount of data has been collected during sports activities thanks to the advances in high-speed cameras and computer vision (CV) techniques.
While this data plays a central role in understanding players' performance and developing winning strategies,
it can be challenging to understand the data without presenting it in its physical context.
Augmented sports videos can present sports data directly in actual scenes through embedded visualizations and video effects, 
communicating insights and explaining player strategies in an intuitive and engaging manner. 
Thus, augmented sports videos have been widely used by TV channels~\cite{espn}, 
fan clubs~\cite{tennisTV}, 
and analysts~\cite{perin2018state, stein2017bring} to present sports data,
influencing billions of sports enthusiasts around the world.

However, creating augmented sports videos is a demanding and time-consuming task~\cite{chen2021}, as it requires skills in areas such as data analysis,
data visualizations, and video editing.
The gap between
the difficulty in creating augmented sports videos
and the strong market demand for augmented sports videos
has spawned very successful commercial systems, such as Viz Libero~\cite{vizrt} and Piero~\cite{piero}.
These systems, however, target expert video editors and require the manipulation of low-level graphical elements. 
This leads to a high entry barrier for sports analysts, who usually focus on presenting analytical insights 
and lack sufficient video editing skills.
Recently, Zhu-Tian et al.~\cite{chen2021} presented VisCommentator, 
a fast prototyping tool for augmenting table tennis videos,
enabling analysts to augment the video 
by interacting with the data of video objects
instead of low-level graphical elements.
However, analysts often express their findings 
as high-level insights, 
such as \quot{Federer hits a backhand down the line}, 
rather than data (\emph{e.g.}, 
the player's position and the ball trajectory). 
Consequently, to visualize an insight, 
an analyst usually needs to ``translate'' the insight into data equivalents
and then map them to visualizations embedded in the videos,
which is tedious and has a heavy cognitive load.

However, the tedious ``translation'' process
implies a latent mapping between the high-level insights and the visualizations.
Such a latent mapping provides an opportunity to facilitate the creation of augmented sports videos
by directly creating visualizations to augment the videos based on the user's insights.
In sports, perhaps the most common way to express insights is using natural language,
\emph{e.g.}, 
commentators give real-time comments on live games,
analysts report their analytical findings of sports videos in oral presentations,
and journalists summarize key events in textual documents.
This work, thus, explores how to leverage natural language to facilitate the creation of augmented sports videos.

This work aims to augment a sports video clip based on a given commentary text.
To achieve this goal, 
we first identify three tasks inspired by existing text-to-visuals systems~\cite{crosscast, Xia2020, Cui2020}:
1) detecting the visualizable entities in the text,
2) mapping these entities into visualizations,
3) scheduling these visualizations to play with the sports video.
To tackle these three tasks,
we collected and analyzed \numClip{} sports video clips, as well as their corresponding commentaries, of six different sports
with three main questions in mind -- 
\emph{What text entities can be visualized (Q1)?};
\emph{How can we visualize these entities (Q2)?}; 
and \emph{How do we schedule these visualizations with the video (Q3)?}
Based on our formative study, we have identified that
four categories of entities, namely, \obj{}, \act{}, \data{}, and \emo{},
in the commentaries can be visualized,
entities in different categories can be visualized by different embedded visualizations, 
and the scheduling of visualizations depends on the style of the commentaries, 
\emph{i.e.}, \emph{analyst} or \emph{play-by-play}, in which the video is paused or not, respectively.

Based on the findings in our formative study, 
we have designed and developed \system{} (\autoref{fig:teaser}), 
a proof-of-concept system that \cmo{takes
textual commentaries,
racket-based sports (\emph{e.g.}, table tennis, tennis) 
videos, 
and sports data (\emph{e.g.}, player and ball positions, key events) as the input} 
to produce augmented sports videos.
\cmo{The inputted sports data can be extracted from the videos by using CV models or manually prepared.}
In contrast to most existing language-driven visualization creation tools,
\system{} sees the text as an information source instead of a command (\eg, \quot{show me the bar chart!}) to create the visualizations.
\system{} features three components, \ie, \emph{Entity Detector}, \emph{Entity Visualizer}, and \emph{Visualization Scheduler}, to complete the aforementioned three tasks.
Behind these three components is a set of state-of-the-art natural language processing (NLP) models.

\system{} is a technique that can be applied in different application scenarios.
To demonstrate the usage of \system{}, 
we exemplify two application scenarios,
including authoring augmented racket-based sports videos using text
and augmenting historical sports videos based on auditory comments.
To evaluate the effectiveness of \system{},
we first conducted a technical evaluation that focuses on the accuracy of the Entity Detector, 
which is the foremost step in the pipeline,
and achieved a good performance of an F1 score of 0.9.
A task-based expert evaluation with eight sports analysts
confirmed the utility, effectiveness, and high satisfaction of our language-driven creation method.
We also discuss promising future opportunities implied by observations and feedback from the study.

In summary, our main contributions are threefold:
First, we conduct a formative study and subsequent identification and discussion of design considerations for augmenting sports videos based on commentary text.
Second, we present the design and implementation of \system{}, a proof-of-concept system that augments racket-based sports videos based on a piece of commentary text.
Third, we demonstrate two exemplar applications based on \system{} and report on a technical evaluation and expert feedback.
\section{Related Work}

\para{Embedded Visualization in Sports Videos.}
The advances in high-speed cameras and CV techniques
have made sports data from videos increasingly available~\cite{perin2018state}.
To present the data in a meaningful context,
researchers have proposed methods that visualize
the data together with the videos, 
such as side-by-side~\cite{DBLP:conf/chi/FuS22} and embedded views~\cite{stein2017bring}. 
This work focuses on embedded visualizations in sports videos.

The idea of embedding sports data in its physical context is not new.
In the early stage,  researchers embedded visualizations in court diagrams, 
which can be seen as a simplification of the real-world scene~\cite{DBLP:journals/cgf/SachaASSKAJ17, DBLP:journals/tvcg/AndrienkoAABBFH21, DBLP:conf/avi/AndrienkoABLW18, DBLP:journals/tvcg/PerinVF13, DBLP:journals/tvcg/WangWCZZW21}.
Recently, 
with more powerful graphics cards and advanced CV techniques,
researchers have started to explore ways to embed visualizations in sports videos directly~\cite{fischer2019video}.
For example, 
Stein et al.~\cite{stein2017bring} introduced a system 
that takes raw footage of soccer games as the input
and automatically visualizes relevant analytic measures of the players in the video.
Stein et al.~\cite{stein2018} further extended their work with a framework that semi-automatically
decides what measures should be presented at a specific moment.
However, most of these research systems were developed for exploration purposes
and thus do not allow users to visualize their insights in the videos for communication purposes.
On the other hand, 
industry companies have developed commercial systems to assist in creating embedded visualizations in sports videos.
For instance, 
Piero~\cite{piero} and Viz Libero~\cite{vizrt} are powerful video editing tools
that have been used for annotating sportscasting and producing TV programs.
CourtVision~\cite{courtvision}, developed by Second Spectrum~\cite{secondspectrum},
is a basketball watching system that automatically embeds players' status information in basketball videos to engage the audience.
However, these industrial products target proficient video editors,
who focus on the manipulations of graphical marks,
leading to complex interface actions and a steep learning curve for sports analysts.

Perhaps the most closely related work to the present research is VisCommentator~\cite{chen2021},
which is an application that enables sports analysts to create augmented sports videos
by selecting \emph{sports data}.
\cmo{
Our work aims to further ease the creation process 
and support users to augment sports videos
by directly expressing \emph{sports insights in natural language}.
Compared to VisCommentator,
our system is a modular technique that provides a higher abstraction level,
leading to a different system design, interactions, and technical implementations.
}

\para{Natural Language Interfaces for Visualization.}
Recent achievements in NLP have reignited interest in 
using natural language interfaces (NLIs) for creating data visualizations.
Compared to traditional visualization creation tools,
systems with NLIs enable users to express their
intentions via natural language rather than interface actions or coding,
thereby lowering the barrier to visualizing data.

Existing NLI systems can be roughly divided into \emph{explicit} and \emph{implicit} approaches.
Explicit NLI systems~\cite{NL4DV, DataTone, Eviza, Srinivasan2021} treat the natural language as commands
and require users to illustrate their intentions explicitly.
For example, DataTone~\cite{DataTone} allows users to create visualizations of their desired data
by typing, \emph{e.g.}, \quot{Show me medals for hockey and skating by country}.
The NL4DV toolkit~\cite{NL4DV} takes a tabular dataset and a text query as the input
and outputs visualizations in the form of JSON specifications.
On the other hand, 
implicit NLI systems
view the text descriptions as another representation of the visual content, 
automatically converting the text to visual content and thus
enabling users to create visual content implicitly.
Extensive research in computer vision, computer graphics, and human-computer interaction
has explored the automatic conversion of descriptive text into visual content,
such as images~\cite{crosscast, Xia2020}, 3D shapes~\cite{Chen2018} and scenes~\cite{Coyne2001, Chang2014}, documents~\cite{crossdata}, and short video clips~\cite{Marwah2017}.
In recent years,
with the development of generative adversarial networks,
a plethora of systems~\cite{text-to-image, text-to-image2, stackgan++, text-to-video, wu2018multimodal} have been proposed to generate visual content based on text descriptions.
\cmo{However, none of those works investigates generating augmented sports videos from textual commentaries.}

Our work employs an implicit approach 
and enables users to create augmented videos by expressing insights in natural language.
In the visualization community,
only few implicit NLI systems for visualization creation exist.
A representative example is Text-to-Viz~\cite{Cui2020},
which extracts semantic information from the user's description of insights 
and maps it into static infographics.
We share a similar spirit but focus on 
creating visualizations from text to augment sports videos,
which poses extra challenges 
as both the text and video need to be considered in the creation process
and the resulting visualizations need to be dynamic.

\para{Auto-Generation Techniques for Data Visualization.}
Creating data visualizations usually involves 
exploring data to discover insights
and mapping them into proper visualizations,
both of which are demanding and time-consuming.
Thus, to ease the creation process, researchers have developed techniques
to automate or semi-automate the data exploration and visual mapping steps.
For example, 
to facilitate the data exploration step, 
DataShot~\cite{WangSZCXMZ20} employs an auto-insight technique to suggest interesting data patterns 
from spreadsheets and generated factsheets.
By using a pattern detection engine,
DataToon~\cite{kim2020}, an authoring tool for data comics, 
automatically suggested salient patterns of the input network data.
To ease the visual mapping step,
prior research has proposed template-based methods 
to automatically map a subset of data to different types of visualizations,
including charts~\cite{voyager}, infographics~\cite{WangSZCXMZ20}, and data animations~\cite{Amini2017}.
The templates in these systems are usually hand-crafted based on prior knowledge and empirical studies.
Research systems, such as DeepEye~\cite{deepEye}, Draco~\cite{draco}, VizML~\cite{viznet}, and AutoTimeline~\cite{chentimeline}, have used machine learning models to automatically learn and extract templates from existing visualizations.

In contrast to these systems, our work focuses on extracting information from natural language rather than structured datasets
and converting it to embedded visualizations for augmenting sports videos.
In this sense, our system is closer to Text-to-Viz~\cite{Cui2020}, 
which generates infographics based on text,
but has a very different output, augmented sports videos, which is more difficult and constrained. 

\para{Immersive Sports Visualization.}
Fundamentally,
augmented sports videos share a similar spirit with immersive sports visualization,
which leverages virtual or augmented reality devices to 
visualize sports data 
in either
simulated~\cite{shimizu2019sports, tanaka2018scope, ShuttleSpace, tsai2020feasibility, TIVEE} or real courts~\cite{ARBasketballTrainning}.
While the immersive visualizations in these systems have proven to be effective for sports data analysis,
most of them are predefined and lack customizability.
Thus, users cannot flexibly create new immersive visualizations to express their intentions.
Our work adds to the direction of immersive sports visualization 
but focuses on helping users create visualizations to present data in the physical context shown in a video.
\section{Text to Augmented Videos: A three-step approach}
\label{sec:formative_study}

\begin{figure}[h]
\hspace*{-1.4cm}
  \centering
  \includegraphics[width=1.15\linewidth]{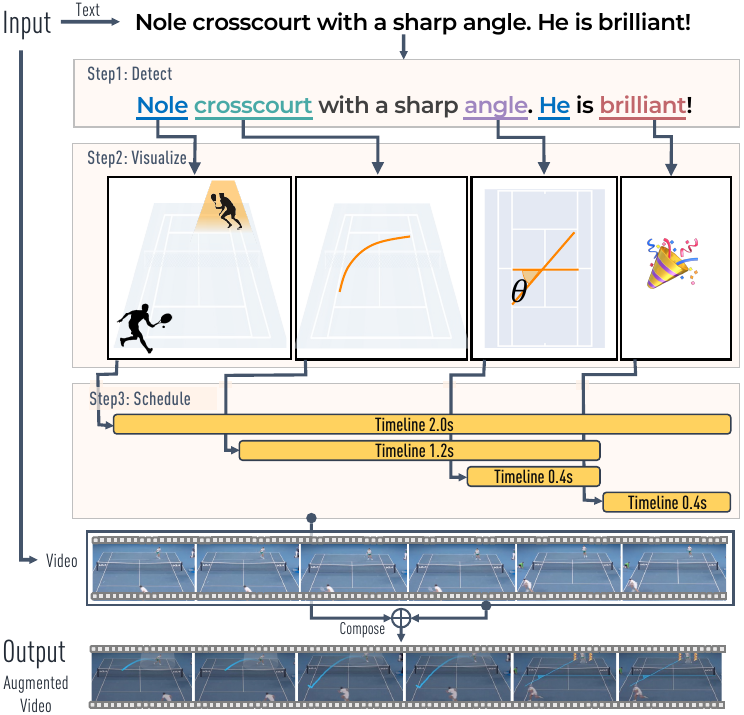}
  \caption{
  A three-step approach to augment sports videos with embedded visualizations based on text commentary.
  The three steps include detecting visualizable entities in the text, mapping them to visualizations, and scheduling the visualizations in the video.}
  \label{fig:framework}
\end{figure}

The goal of this work is to augment a sports video clip based on a given commentary text
by automatically converting the text into embedded visualizations.
To achieve this goal,
inspired by existing text-to-visuals systems~\cite{crosscast, Xia2020, Cui2020},
we propose a three-step approach (\autoref{fig:framework}) 
that decomposes the problem into three tasks:
1) detecting the visualizable entities in the text,
2) mapping the entities to visualizations, and
3) scheduling the visualizations to play with the video.
These tasks lead us to the following three questions:
Q1--\emph{What text entities can be visualized?};
Q2--\emph{How can we visualize these entities?};
Q3--\emph{How do we schedule these visualizations with the video?}
To understand these questions,
we conducted a formative study 
by collecting and analyzing \numClip{} sports video clips
and their accompanying commentaries.

\subsection{Data Collection and Analysis}
There are many publicly available
text sources that comment on sports videos,
including sports commentaries, 
game-related reports, articles, and open discussions (\eg, posts in online forums).
In this work, 
we decided to collect sports commentaries 
since they are usually given 
by sports experts and contain rich insights.

\begin{figure*}[tbh]
  \centering
  \includegraphics[width=\linewidth]{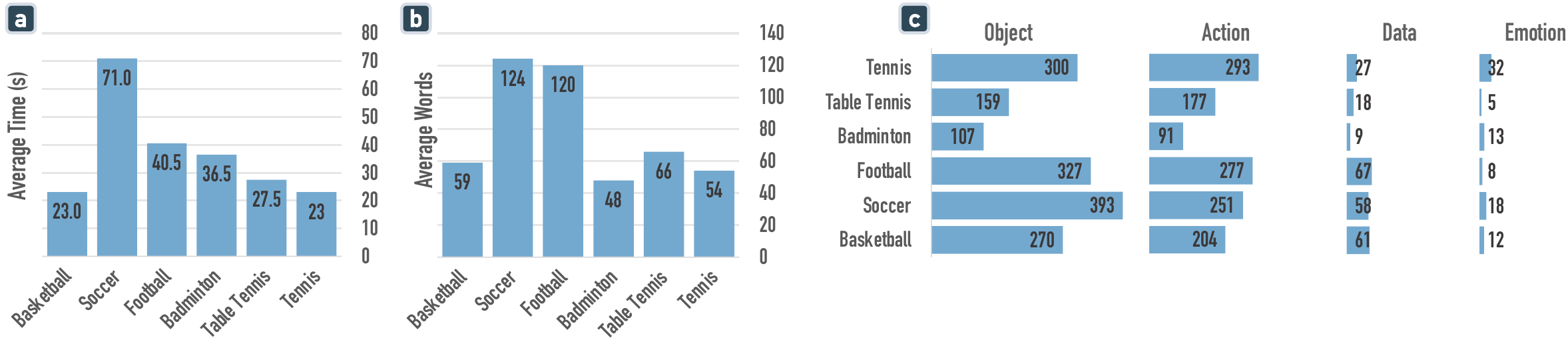}
  \vspace{-4mm}
  \caption{The average a) duration of videos and b) number of words of commentaries per sport in the collected dataset. c) The number of entities per category in different sports.}
  \vspace{-2mm}
  \label{fig:statistics}
\end{figure*}

\para{Collection.}
Following the methodology in~\cite{chen2021},
we harvested a collection of commentaries that
cover three team-based sports (\ie, basketball, soccer, and American football) 
and three racket-based sports (\ie, tennis, badminton, and table tennis) from the internet.
Specifically, we searched sports videos with English audio commentaries
from Google Videos by using the keywords ``SPORT + full match'', 
where SPORT is one of the six ball sports.
We downloaded five videos for each sport from the top query results,
totally gathering \numVideo{} sports videos lasting over \numMinutes{} minutes. 
Our collection considered both the quality (\ie, millions of views) 
and diversity (\ie, from various TV channels, including member-only ones such as ESPN+). 
Most of the videos were final games of famous sports events, 
such as the Olympic Games, FIFA world cups, NBA games, Grand Slams, and the Super Bowl.
Note that our unit of analysis was not an entire game 
but a specific meaningful moment in the game (\eg, a goal, a rally).
Thus, for each video, we manually sampled at least four clips 
following two criteria:
1) the clip should cover a highlighted moment (curated by TV channels) of the game;
2) the commentaries of this clip should be closely related to the sports event happening in the clip
(\ie, commentaries about player anecdotes were thus excluded).
Finally, our dataset included \numClip{} clips lasting 92 minutes,
which aligns with similar visualization research~\cite{amini2015, DBLP:journals/tvcg/ShuWTBWQ21,Thompson2020}.
All videos were transcribed manually by native English speakers, resulting in 12545 words.
Figure~\ref{fig:statistics}a and b show the average duration of the videos and the average number of words in the commentaries of different sports.

\para{Analysis.}
We analyzed the dataset both qualitatively and quantitatively.
Three of the authors first followed an open coding process to
analyze the \numClip{} clips independently, with the three questions (Q1-3) in mind.
The codes were then refined through multiple rounds of discussions with other co-authors.
The investigation also referred to prior research on 
text-driven visual content generation~\cite{Cui2020, crosscast},
sports visualizations~\cite{perin2018state}, 
data-driven videos and animated graphics~\cite{amini2015, DBLP:journals/tvcg/ShuWTBWQ21,Thompson2020},
and augmented sports videos~\cite{chen2021}.
Findings revealed that there are four categories of entities that can be visualized in the text (Q1),
entities in different categories can be visualized with different embedded visualizations (Q2),
and these visualizations can be scheduled with the video in two different ways (Q3).
We further conducted a quantitative analysis
to count the occurrences of the four categories of entities in the dataset (\autoref{fig:statistics}c). 
The findings are detailed in the following sections.

\subsection{Q1: What Text Entities Can Be Visualized?}
\label{ssec:Q1}

Different from traditional visualization systems 
that generate visualizations based on structural data (\eg, spreadsheets),
our goal is to generate visualizations from sports commentaries,
which can be unstructured, messy, and full of uncertainty.
Thus, the first step to achieving our goal is to recognize the
visualizable entities in the text.
According to our analysis of the dataset,
we identified four categories of visualizable entities (highlighted in \texttt{typewriter} font),
which are introduced from concrete, objective to abstract, subjective:

\begin{itemize}
    \item[\textbf{\texttt{E1}}] \textbf{\Objs{}} are physical entities that can be seen in the video,
    which usually serve as the referent for other visualizations.
    Among all \objs{},
    we found two kinds of \objs{} were mentioned most frequently,
    namely, \emph{players} (68.83\%) and 
    \emph{places} in the court (10.73\%). 
    Players were usually mentioned by their names 
    and pronouns
    .
    All \objs{} are noun words in the commentaries.
    
    \item[\textbf{\texttt{E2}}] \textbf{\Acts{}} are performed by \objs{} and can also be seen in the video.
    In the dataset, 
    we found that \acts{} can be roughly divided into
    \emph{sports-general} (80.27\%) 
    and \emph{sports-specific} (19.63\%).
    Sports-general \acts{}, such as \emph{hit}, \emph{run}, and \emph{cover}, 
    are usually verbs and exist in all the six ball sports.
    In contrast, sports-specific \acts{} are terms used in specific sports,
    such as \emph{down the line} in tennis, \emph{pick and roll} in basketball.
    Domain-specific \acts{} can be nouns (\eg, \emph{crosscourt}) or adjectives (\eg, \emph{backhand}), 
    both of which can be used as verbs in the commentary, 
    such as \emph{``Federer backhand on the run,''} \emph{``Djokovic down the line.''}

    \item[\textbf{\texttt{E3}}] \textbf{\Data{}} is usually generated by \acts{} and cannot be seen in the video.
    Prior research~\cite{perin2018state} categorized \data{} in sports videos into 
    \emph{tracking data}, which is inherently spatial and temporal,
    and \emph{non-tracking data}, which is rather abstract.
    We also found these two kinds of data were brought up in the commentaries (16\% and 84\% for tracking and non-tracking data, respectively).
    \Data{} in the commentaries usually are noun words or numbers,
    such as \emph{speed}, \emph{5 meters}, \emph{winning rate}, and \emph{won 64\%}.

    \item[\textbf{\texttt{E4}}] \textbf{\Emos{}} are the subjective feeling of a game 
    expressed by the commentators, adding to the exciting atmosphere of the game.
    \texttt{Emotional cues} cannot be seen in the video but can be felt in the game and the commentaries.
    In the dataset, \emos{} can be adjectives (\eg, \quot{Phenomenal!}), interjections (\eg, \quot{Wow!}), or analogies (\eg, \quot{make it just like Quidditch.})
\end{itemize}

\para{Relationships Among Entities.}
In a commentary,
the visualizable entities are usually not isolated -- instead, they are inherently connected.
As discussed above, an \obj{} can perform an \act{} that generates \data{}.
We noticed that such kinds of relationships are embedded in the linguistic structures of the natural language, 
which can, and should, be extracted and leveraged to specify the visualizations.
For example, by identifying the subject (\ie, an \obj{}) of an \act{}, 
we can visualize the generated tracking data
in the correct position in the video.

\subsection{Q2: How to Visualize the Entities in the Video?}
\label{ssec:Q2}

After recognizing the visualizable entities in the text,
the next step is to map them into proper visualizations
that can be embedded in the video.
\cmo{
While prior work~\cite{chen2021} studied the visualizations of sports data (\ie, tracking or non-tracking) in videos, 
how to visualize commentary entities in videos remains unclear.
Based on our analysis of the commentaries and the videos, 
multiple rounds of discussion and iterations with
a domain expert (a sports science professor who provides data analysis and consultancy services for national sports teams)},
and the previous research on video-based sports visualizations,
we propose the following methods to visualize each category of entities:

\begin{itemize}
    \item \textbf{\Obj{}} entities can be directly seen in the video.
    Thus, to visualize \obj{} entities,
    we can directly \emph{highlight} the corresponding objects in the video.
    Various types of \objs{} can be highlighted differently.
    For example, we can use spotlights and rectangle marks to highlight players and places on the court, respectively. 

    \item \textbf{\Act{}} entities can also be seen in the video,
    but they do not have a physical existence.
    On the other hand, \acts{} are performed by \objs{} and generate \data{}.
    Hence, to visualize an \act{} entity, 
    we can \emph{highlight} the \obj{} when she/he/it is performing the \act{}
    or \emph{visualize} the \data{} generated by the \act{}.
    Compared to highlighting the subject of an \act{},
    visualizing the invisible \data{} of the \act{} can reveal more insights and better engage the audience.
    However, one must know what \data{} is generated by the \act{} to achieve the visualization,
    which is relatively easy for sports-general \acts{} but can be challenging for sports-specific ones.

    \item \textbf{\Data{}} entities, according to prior research~\cite{chen2021}, 
    can be visualized in the video by mapping them to different visual representations based on their types.
    Specifically, 
    we can naturally \emph{embed} tracking data into the video as it is always associated with a specific space and time in the video.
    For non-tracking data, we can \emph{annotate} the video with labels to show it.

    \item \textbf{\Emos{}} are the most abstract entities and are sometimes not directly associated with
    the objects in the video. 
    To visualize \emo{} entities, 
    a straightforward way is to \emph{display} semantic-related pictograms, such as emojis, in the video. 
    While other more advanced methods are possible, 
    we consider exploring them as beyond the scope of this work,
    as the research of affective visualization is still in its infancy~\cite{DBLP:journals/tvcg/LanSWJC22}.
\end{itemize}

In summary, 
entities in different categories can be visualized differently,
such as highlighting the entities in the video (\obj{}, \act{}),
visualizing the data generated by the entities (\act{}),
embedding or annotating the entities into the video (\data{}),
or presenting the entities using emojis (\emo{}).
All these methods map the entities into visualizations embedded in the video.
To select the specific visualizations,
we follow previous work~\cite{chen2021} that summarized a design space of embedded visualizations in augmented sports videos.

\subsection{Q3: How to Schedule the Visualizations in the Video?}
\label{ssec:Q3}

After mapping the entities into visualizations,
the last step is to schedule the visualizations in the video. 
This step entails deciding \emph{when} to display a visualization and \emph{how long} to display it for.
This, intuitively, depends on several factors, including the text, the video, and the visualizations themselves.
In our analysis, we noticed that this question is particularly
related to the commentary style of the text.
Specifically, there are two major types of sports commentators --
\emph{play-by-play} commentators,
who need to articulate each play and event of an often fast-moving sports game,
and \emph{analyst} commentators, 
who provide expert analysis and background information of the game.
These two types of commentaries lead to two different rendering methods,
wherein the visualizations are scheduled differently:

\begin{itemize}
    
    \item \textbf{Play-by-play} mode renders the visualizations without pausing the video,
    since the visualizations are generated by commentaries that describe
    the ongoing content of the video.
    In this mode, the scheduling of the visualizations depends on both the text and the video.
    For example, 
    the visualization of an \act{} should be displayed when it is mentioned in the text and disappear when the \act{} is finished in the video. 
    
    \item \textbf{Analyst} mode usually renders the visualizations by pausing the video,
    since the commentaries are often given during a break or replay of the game
    and contain too much information to be visualized in a short moment.
    In this mode,
    the scheduling of the visualizations only depends on the text
    since the video is paused.
    Thus, the start time and duration of a visualization
    should be decided by when and how it is mentioned in the text.
\end{itemize}

While we divide the rendering and scheduling methods of visualizations into two types 
based on the commentary styles, 
they are not meant to be the only solution.
For example, an analyst-style commentary may also be able to be rendered without
pausing the video.
We leave the comprehensive exploration of the rendering and scheduling of visualizations in augmented sports videos for future research.
\begin{figure*}[bth]
  \centering
  \includegraphics[width=\linewidth]{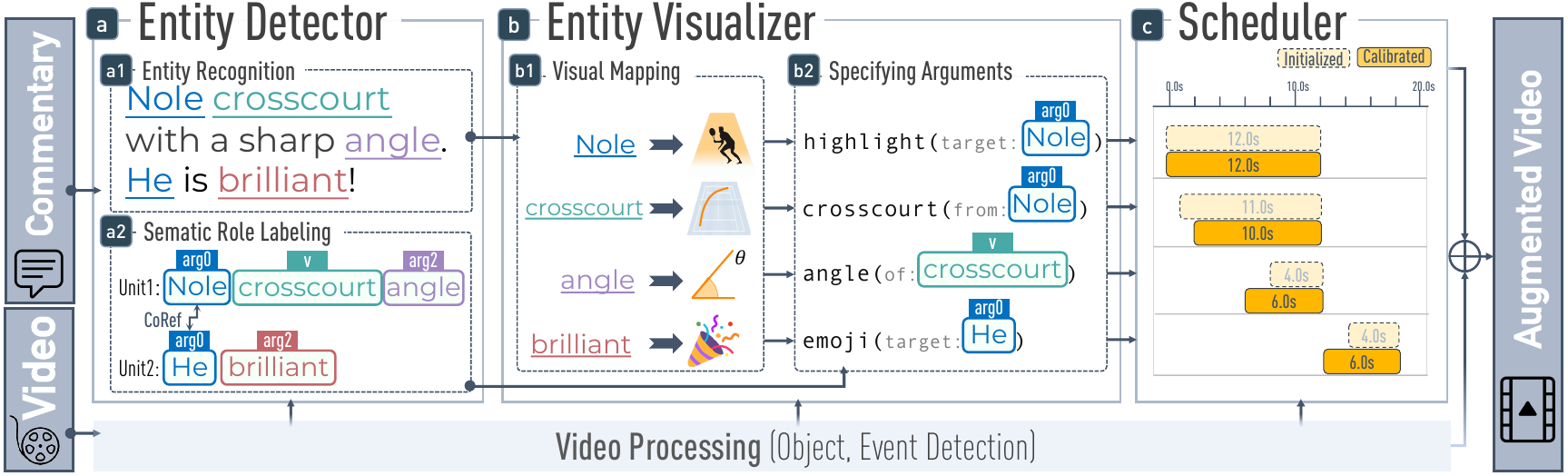}
  \caption{
  \system{} detects the visualizable entities in the text (a1)
  and groups them into semantic units (a2).
  Next,  the entities are mapped to visualizations (b1) with arguments specified by the semantic units (b2). 
  Finally, the system initializes and calibrates the schedules of the visualizations based on the reading time of the text and the video events (c). All three steps are built upon the video processing components.
  }
  \vspace{-4mm}
  \label{fig:architecture}
\end{figure*}

\section{\system{}: System Design and Implementation}

To realize the three-step approach, 
we design and implement \emph{\system{}}, 
a proof-of-concept system that
creates augmented videos for racket-based sports. 
\system{} consists of three major components -- 
\emph{Entity Detector},
\emph{Entity Visualizer}, 
and \emph{Visualization Scheduler}
-- for each step in the approach, respectively.
It takes a piece of text, a video clip\cmo{, and sports data extracted from the video clip as the input},
and outputs an augmented video.
Figure~\ref{fig:architecture} displays the pipeline of \system{}.
We first introduce the video processing and rendering techniques \system{} is built upon,
followed by details of the three components.

\subsection{Video \cmo{Data Extraction} and Rendering}
While this research particularly focuses on leveraging 
natural language to create augmented sports videos,
the implementation of \system{} is built based on advanced CV techniques.
We follow previous work~\cite{chen2021} 
and use several machine learning models
to extract data from the videos.
Specifically,
we use Detectron2~\cite{wu2019detectron2}, a detection and segmentation platform
that features a mask-rcnn~\cite{he2017mask} with 
an ImageNet~\cite{Deng2009} pre-trained RestNet-50~\cite{resnet} as the backbone
to detect
the bounding boxes of the players and ball, 
the skeletons of the players, and the court lines. 
We also use TTNet~\cite{ttnet} to detect ball events, including ball bounce and net hit.
For player events,
we utilize the distance between the ball and the player's right hand to detect stroke events.
The extracted data is used in the following steps for creating visualizations to augment the videos. 
We refer readers to~\cite{chen2021} for the technical details.

To render visualizations in the videos,
we further need to know 
the camera parameters
and the player identities (\ie, who is far from or near the camera).
While the camera parameters can be obtained 
using camera calibration techniques~\cite{DBLP:conf/spieSR/FarinKWE04, DBLP:journals/pami/Zhang00} on
the court lines detected from the video,
we treat them and the player identities as known meta information
in the current implementation.

\subsection{Entity Detector}
The first step of the framework is to detect the visualizable entities in the text (\autoref{fig:architecture}a1),
\ie, \objs{}, \acts{}, \data{}, and \emos{}.
Additionally, we need to extract their relationships and group them into semantic meaningful units (\autoref{fig:architecture}a2).
To this end,
we leverage a series of state-of-the-art NLP techniques to process the input text:

\para{Detecting Visualizable Entities.}
Detecting entities in a piece of text is a fundamental task in NLP called Name Entity Recognition (NER)~\cite{nadeau2007survey},
which locates and classifies segments in the text into predefined categories, such as person, location, organization, \emph{etc}.
To detect the entities in a sentence, three steps are taken: 
1) tokenizing the sentence into a word sequence,
2) converting the tokens into feature vectors,
and 3) classifying the feature vector of each token into categories.
We achieved these three steps by using Spacy~\cite{spacy}, 
an industrial-strength NLP toolkit that provides pre-trained transformer-based~\cite{devlin-etal-2019-bert} language models to tokenize, featurize, and recognize entities in a sentence.
To improve the recognition performance of Spacy, 
we fine-tuned its pre-trained model with the commentary examples we collected 
and \cmo{extended its 
pattern matching step with sports glossaries collected from Wikipedia~\cite{tGlo, ttGlo}}.

We found that some expressions, 
such as \emph{he} and \emph{here}, refer to other entities in the text and need to be resolved to be able to visualize them.
Thus, to solve this issue, 
we employed a neural co-reference resolution model~\cite{neuralCoreference},
which can find all expressions that refer to the same entity in a text.
In sum, 
the output of this step is a list of entities with their categories 
and pointers to their referents if they exist.

\para{Grouping the Entities into Semantic Units.}
As discussed in Sec.~\ref{ssec:Q1}, 
the entities in the text are usually connected together at the semantic level, \eg, 
an \obj{} performs an \act{} that generates \data{}.
Such semantic relationships are critical to specify the visualizations.
Take \quot{Federer hits the ball to the backhand court} as an example, 
in which four entities are detected, \ie, \texttt{Federer}, \texttt{hits}, \texttt{ball}, and \texttt{backhand court}.
We cannot visualize \texttt{hits} and \texttt{backhand court} without knowing the subject and possessive noun.
To extract the semantic relationships among the entities,
we leverage the Semantic Role Labeling (SRL)~\cite{palmer2010semantic} technique,
which detects the latent predicate-argument structure of a sentence
and classifies the roles of each word.
For example, \texttt{hits} will be detected as a predicate with \texttt{Federer}, \texttt{ball}, and \texttt{backhand court} 
as argument 0, 1, and 2, respectively.
With this structure, 
we can easily group the entities into units like \emph{who} (argument 0) \emph{did what} (predicate) to \emph{whom} (argument 1) in \emph{what ways} (argument 2).
In our implementation, we use a BERT-based model~\cite{DBLP:journals/corr/abs-1904-05255} to detect and classify the semantic roles of the entities.
The output of this step is a list of units organized in the predicate-argument structure.

\subsection{Entity Visualizer}
The next step is to visualize the detected entities,
which we achieve by 
mapping the entities to visualizations (\autoref{fig:architecture}b1)
and specifying the visualizations' arguments based on the text (\autoref{fig:architecture}b2):

\para{Mapping the Entities to Visualizations.}
As discussed in Sec.~\ref{ssec:Q2}, 
different categories of entities can be mapped to different visualizations.
To achieve the visual mappings,
we developed a dictionary based on a design space of augmented sports videos~\cite{chen2021} 
and an emoji searching engine.
For example,
players (\obj{}) are mapped to spotlight highlight effects; 
ball angles (\data{}) are mapped to embedded visualizations in the court;
\quot{brilliant} (\emo{}) is mapped to a celebration emoji.
More details can be found in the supplemental material.

A particular challenge is to map \acts{} to visualizations,
especially for the sports-specific \acts{}, such as \emph{crosscourt}, 
which are abstract and usually require case-by-case designs.
Two steps are thus taken to tackle this challenge:

\begin{enumerate}[noitemsep,topsep=0pt]
    \item \underline{Mapping sports-general \acts{} to tracking-data}:
    In the dictionary, we manually specify the mappings from sports-general \acts{}
    to the tracking data they generate, \ie,
    \texttt{run} and \texttt{hit} are mapped to the \texttt{player trajectory}
    and \texttt{ball trajectory}, respectively.
    The mappings are initialized with 21 sports-general \acts{} 
    and then extended with their synonyms using word embedding~\cite{spacy}, 
    \eg, \texttt{hit} is extended with \texttt{stroke}, \texttt{shoot}, \emph{etc}.

    \item \underline{Mapping sports-specific \acts{} to sports-general ones}:
    Sports-specific \acts{} are terminologies that are difficult to be extended from other \acts{} using synonyms matching.
    To map sports-specific \acts{} to the tracking data in a generalizable way,
    we leverage sports glossaries that explain these terminologies in plain text. 
    For instance, 
    \texttt{crosscourt} is explained as \emph{``Hitting the ball into the diagonal court''} in a tennis glossary~\cite{tennisGlo}.
    Based on this explanation, 
    we can use synonym matching to map \texttt{crosscourt}
    to \texttt{hit} and then map to \texttt{ball trajectory} in our dictionary.
\end{enumerate}

With these two steps,
we can map the \acts{} to the tracking data they generated,
which will be mapped to embedded visualizations.

\begin{figure}[th]
  \centering
  \includegraphics[width=0.8\linewidth]{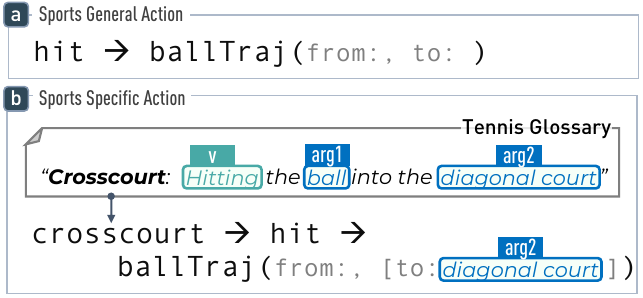}
  \caption{
    a) The visualization of \texttt{hit} is manually specified, which takes two arguments, \ie, \emph{from} and \emph{to}.
    b) The visualization of \texttt{crosscourt} can be generated based on its text explanation in the tennis glossary, 
    which is a variant of \texttt{hit} with a default argument, \texttt{diagonal court}.
  }
  \label{fig:dict}
\end{figure}

\para{Specifying the Arguments of the Visualizations.}
To visualize the entities in the video,
we further need to specify the arguments of the visualizations,
which can be extracted from the text based on the semantic relationships among entities.
Take \quot{Federer hits the ball to the backhand court} as an example.
The visualization of the \texttt{hit} action needs two arguments, \ie, \emph{from} and \emph{to} (\autoref{fig:dict}a),
which can be the argument 0 (\eg, \texttt{Federer}) and argument 2 (\eg, \texttt{backhand court}) in the text.
Some sports-specific \acts{} can infer some arguments for their visualizations based on the text explanation.
For example,
by applying the SRL technique to the text explanation of \texttt{crosscourt} (\autoref{fig:dict}b),
we can set the \texttt{diagonal court} as the default value for the argument \emph{to}.
In this sense, the visualization of \texttt{crosscourt} can be seen as 
the one of \texttt{hit} with a default argument, \texttt{diagonal court}.
More details of visualization arguments are in the supplemental material.

\subsection{Visualization Scheduler}

\begin{figure}[h]
  \centering
  \includegraphics[width=1.02\linewidth]{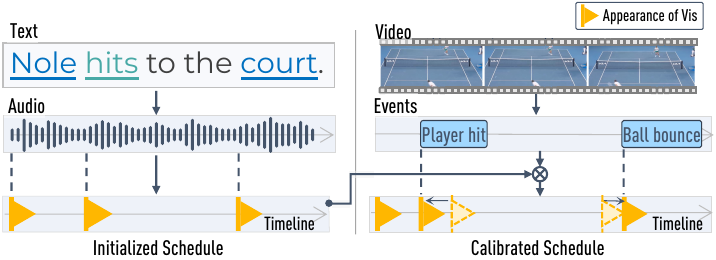}
  \caption{
  Left: The text is converted into audio that initializes the appearance time of each visualization. This initialized schedule can be used to render the visualizations in analyst mode.
  Right: When rendering in play-by-play mode, the appearance times of some visualizations are further calibrated based on video events.}
  \label{fig:scheduler}
\end{figure}

Lastly,
to embed the visualizations into the video,
we need to decide when a visualization should appear and disappear in the video.
We initialize the time schedules of the visualizations based on the text (\autoref{fig:scheduler} left).
The initialized schedules can be used to render the visualizations in analyst mode.
When rendering in play-by-play mode,
we further calibrate the schedules based on the video events (\autoref{fig:scheduler} right).

\para{Initialization.}
Based on our analysis in Sec.~\ref{ssec:Q3}, 
we use the text to initialize the schedules of the visualizations.
Intuitively, a visualization should be displayed when its corresponding entity is read in the sentence.
\cmo{Thus, we employed a text-to-speech neural network~\cite{shen2018natural}
to generate natural speech audio for the input sentence.
With the speech audio, 
we can obtain the start reading time of each entity, 
which is used as the appearance time for the visualization of the entity.
Next, visualizations within the same semantic unit are scheduled to 
disappear at the end of reading the last word of the unit to emphasize their connections.
In analyst mode (\ie, when visualizations are shown as an inserted animation while pausing the video), 
this initial scheduling is sufficient. 
However, when rendering in play-by-play mode, the scheduling needs to be further calibrated.}

\para{Calibration.}
When rendering in play-by-play mode,
the visualizations of \act{} and its argument entities must match the corresponding events.
For example, the visualization of \texttt{hit} 
and \texttt{court} (\autoref{fig:scheduler} left)
should only appear when the player hits the ball and the ball touches the ground, respectively.
\cmo{Thus, we calibrate the schedules of \acts{} and their arguments based on the events detected in the video (\autoref{fig:scheduler} right).
Specifically, for each \act{} and its argument entities,
we look up the corresponding event in the video
by examining the type and time interval of the video events.
If a corresponding event is found,
we use the start time of the event 
as the appearance time for the visualization.
Nevertheless, the schedules could still be sub-optimal, which should be manually refined by the users via external validity.}

\subsection{\cmo{External Validity}}
\cmo{
As an intelligent system, 
\system{} inevitably might derive sub-optimal results
due to the imperfect underlying machine learning models and the limited mapping dictionary.
To support error recovery and visualization personalization,
several methods for external validity can be introduced to the system.
First, 
we can leverage the text entities as an representation to allow users to modify the system outputs.
Specifically, 
the users can select an entity and open a context menu 
to modify its corresponded visualization, as well as the arguments and time schedule of the visualization.
Second, the dictionary of Entity Visualizer should be configurable so that users can modify the mappings persistently. 
Last, the time schedules of visualizations can be visualized along with the video timeline,
enabling users to understand and adjust the schedules.
}

\section{Application Scenarios}
\label{sec:applications}

\system{} is a technique that can be employed in different scenarios
to augment racket-based sports videos.
In this work, 
we implemented two application prototypes
to exemplify the usage of \system{}.

\begin{figure*}[tb]
  \centering
  \includegraphics[width=\linewidth]{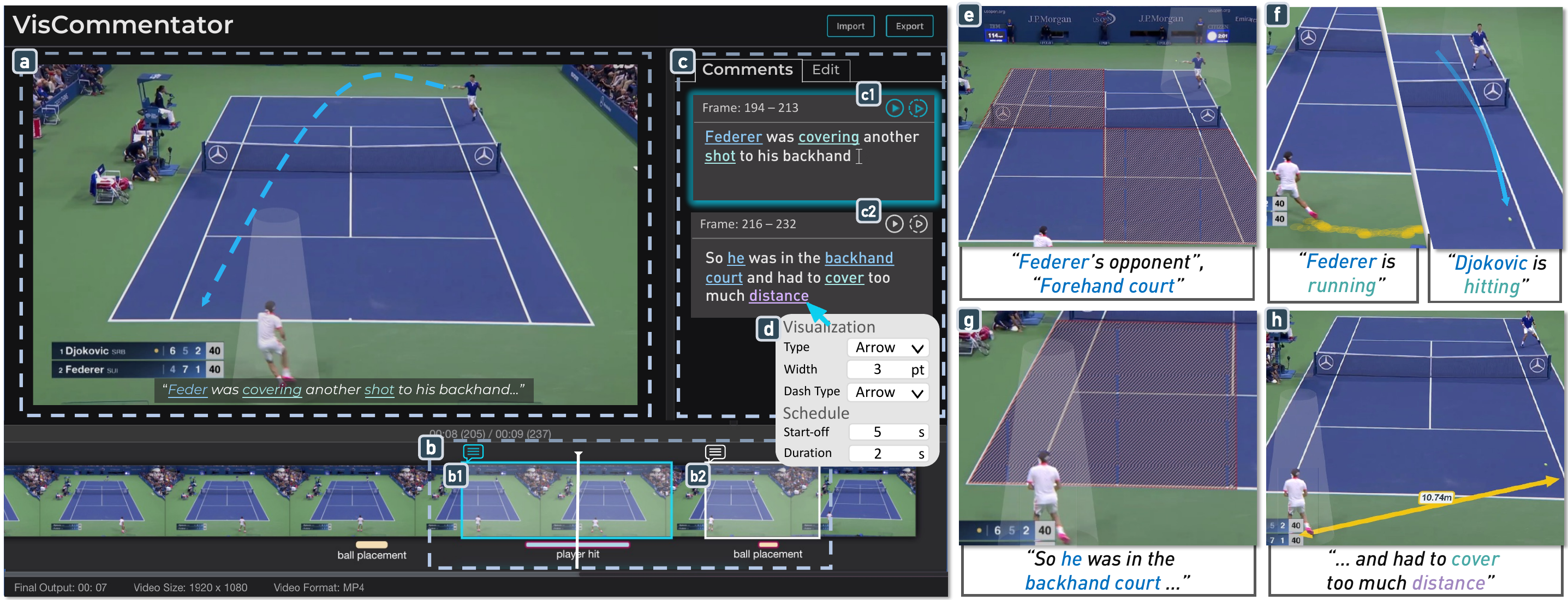}
  \caption{The user interface of VisCommentator after being integrated with \system{}.
  Users can brush on the timeline and comment on the brushed period to e) highlight the players or places, f) visualize tracking data,
  and g) - h) explain insights. \cmo{Users can also d) assess and modify the visualization and schedules of the text entities.}}
  \label{fig:appa}
  \vspace{-4mm}
\end{figure*}

\begin{figure*}[b]
  \centering
  \includegraphics[width=\linewidth]{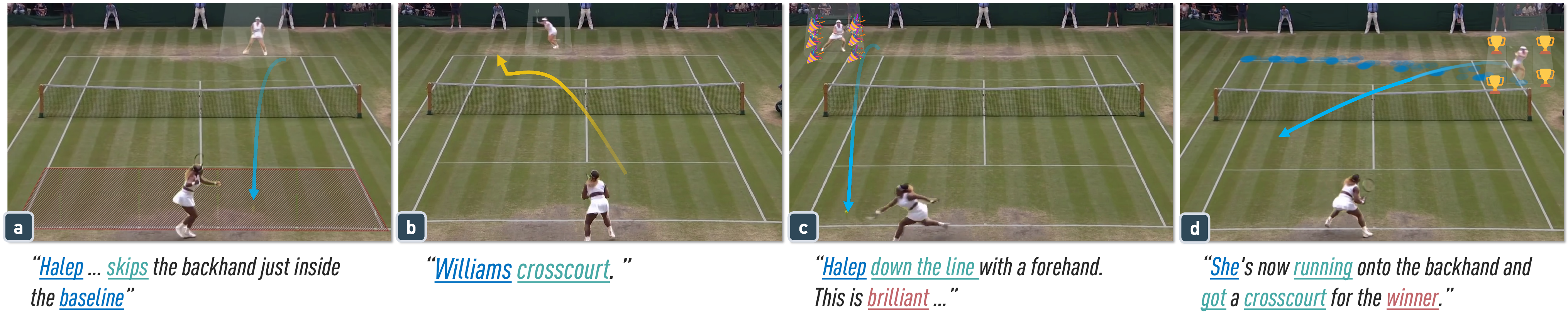}
  \vspace{-5mm}
  \caption{\system{} leverages audio commentaries to augment historical sports videos with embedded visualizations.
  }
   \vspace{-2mm}
  \label{fig:appb}
\end{figure*}

\subsection{Application I: Authoring Augmented Sports Videos}

VisCommentator~\cite{chen2021} is an authoring tool for creating 
racket-based augmented sports videos.
We integrated \system{} into VisCommentator as a sub-system
to 
enable analysts to create augmented sports videos by directly expressing high-level insights in natural language.

\cmo{
To achieve the integration,
we modify the user interface (UI) of VisCommentator to allow text input
and connect its data extractor and renderer to \system{}.
Specifically,
when a user brushes on the timeline (\autoref{fig:appa}b1),
a text input field will show up on the right panel (\autoref{fig:appa}c).
The user can then type in the input field to comment on the video.
Once the user presses the play button,
the text and the data extracted from the video will be passed to \system{},
which generates and schedules embedded visualizations through the three components (\autoref{fig:architecture}).
The scheduled embedded visualizations will be rendered into the video by the renderer of VisCommentator.
}

\cmo{
Additionally, we also provide UI to support external validity of \system{}.
The text entities visualized in the video will be highlighted in the text input field (\autoref{fig:appa}c). 
Users can right click on an entity to assess and modify how it is visualized.
For example, a user can configure the visualization and schedule of the entity \texttt{distance} in~\autoref{fig:appa}d.
Users can also manually create or remove an entity to create their own visualizations if the detection is incorrect,
and switch to the ``edit'' panel to modify other settings such as the visual mapping dictionary.
}

\cmo{
After integrating with \system{}, VisCommentator enables users to create augmented sports videos more efficiently. 
For example, users can simply highlight objects or actions in the video through text (\autoref{fig:appa}d and e).
Besides, users can comment on a moment using a more complex sentence (\autoref{fig:teaser})
or on multiple moments (\autoref{fig:appa}b1 and b2) to create a series of augmented clips (\autoref{fig:appa}a, f, and g).
}

\subsection{Application II: Augmenting Archive Sports Videos}
Many sports events, such as the Wimbledon Championships, 
are broadcast on TV,
recorded as videos with audio commentaries,
and released on online platforms (\eg, YouTube).
While these videos are widely used for analysis or entertainment purposes,
the audio commentaries are usually not fully exploited.
A promising use case is to leverage the audio commentaries to generate visualizations to augment the video,
thereby facilitating the analysis of the games 
and increasing the engagement of watching experiences.

To augment racket-based sports videos by leveraging the audio commentaries,
we implemented an examplar application that integrates a speech-to-text (STT) neural network with \system{}.
Specifically, the application takes a video clip with audio commentaries as the input, 
separates the audio track from the video,
converts the audio commentaries into text by using Silero~\cite{Silero_Models}, an enterprise-grade pre-trained STT model,
and finally generates the augmented video by using \system{}.

Figure~\ref{fig:appb} presents an example produced by the application.
The input video is a BBC sports clip of the women's final in the 2019 Wimbledon Championships.
In the generated augmented video, 
the ball trajectory and the court inside the baseline are visualized
when the commentator describes that 
Halep (\ie, the player far from the camera)
hits the ball \quot{just inside the baseline} (\autoref{fig:appb}a).
Next, the ball trajectory is visualized in yellow color to represent \quot{Williams crosscourt.} (\autoref{fig:appb}b)
Right after Halep's forehand down the line,
the commentator complimented that \quot{this is brilliant}, which is visualized as several celebration emojis around Halep in the augmented video (\autoref{fig:appb}c).
Finally, the augmented video displays Halep's running trajectory followed by the ball trajectory and trophy emojis (\autoref{fig:appb}d)
when the commentator said \quot{She's now running onto the backhand and got a crosscourt for the winner.}

\section{Evaluation}
This section reports technical and expert evaluations on \system{}.

\vspace{-0.5mm}
\subsection{Technical Evaluation}

We evaluated the accuracy of recognizing the four categories of text entities (\autoref{fig:architecture}a1),
which is the foremost step of our approach.
We did not technically evaluate other components because they are either off-the-shelf components with high accuracy (\eg, the SRL model achieves a 0.86 F1-score on benchmark datasets)
or the ground truth is not available (\eg, the visualizations and schedules).
Instead, we conducted expert interviews to collect qualitative feedback on the overall system.

\vspace{-0.5mm}
\para{Dataset.}
We prepared a dataset for entity recognition 
by labeling the entities in each sentence of our collected commentaries (\autoref{fig:statistics}c).
Specifically, 
we \cmo{manually} labeled the start and end index of each entity as well as its category according to our analysis in Sec.~\ref{sec:formative_study}.
The label of each entity was represented as $(start, end, category)$.
We randomly split the dataset into 10 participants for 10-fold cross-validation.

\vspace{-0.5mm}
\para{Model.}
To recognize the entities in a sentence,
we use our dataset 
to fine-tune the pre-trained \textit{en\_core\_web\_trf}~\cite{trf} model provided by Spacy~\cite{spacy}.
The \textit{en\_core\_web\_trf} model is trained on written text such as blogs, news, comments
using the transformer structure~\cite{devlin-etal-2019-bert}.
\cmo{We also extend its pattern matching step~\cite{spacyMatching} with sports glossaries.}
For each round of cross-validation,
we use nine partitions to fine-tune the model and use
the remaining one for the testing.

\begin{table}[h]
	\centering
	\vspace{-2mm}
	\caption{Precision, recall, and F1-score of the entity recognition.}
	\label{table:ent_rec}
	\begin{tabular}{lcccc}
		\toprule
		 \textbf{Entity} & \textbf{Object} & \textbf{Action} & \textbf{Data} & \textbf{Emotion Cue} \\ \midrule
		 Precision & 0.92  &  0.90    & 0.86 & 0.84 \\ 
		 Recall   & 0.95   &  0.95    & 0.90 & 0.91  \\ 
 		 F1-Score & 0.93   &  0.92    & 0.88  & 0.88  \\ 
		\bottomrule
	\end{tabular}
	\vspace{-2mm}
\end{table}

\para{Results.}
Table.~\ref{table:ent_rec} shows the mean precision, recall, and F1-scores of the 10-fold cross-validation.
Overall, the model achieves high accuracy across the four categories.
The accuracy of \obj{} and \act{} 
is comparatively higher than those of \data{} and \emo{}
since the latter have fewer data points.
\cmo{Note that the model, as well as its pattern matching, is data-driven, 
which means our Entity Detector can be improved by
and generalized to other larger datasets.}

\subsection{Expert Evaluation}
To assess the utility and effectiveness of \system{},
we used VisCommentator as a technology probe to conduct a qualitative expert evaluation. 
The study aimed to 
evaluate whether sports analysts can create augmented videos with our system,
observe their creation process, reflect on future improvements,
and collect feedback about the three-step approach and 
how language-driven authoring can facilitate their overall workflows of presenting analytical findings. 

\para{Participants}:
We recruited 8 sports analysts (P1-P8; 3 female; age: 20-58)
from a university sports science department.
All experts majored in Sports Training
with proficient experience in analyzing racket-based sports matches.
P1-P4 particularly focused on analyzing tennis matches,
while P5-P8 focused on table tennis.
P8 was a senior sports analyst lead who
had more than ten years of experience in providing consulting services
for national sports teams.
All the experts only had experience with lightweight video editors, \eg, Tiktok~\cite{tiktok}, 
rather than advanced video editing tools such as Adobe Premiere.
Each participant received a gift card worth \$16 at the beginning of the session, 
independent of their performance.

\para{Tasks}:
The participants were asked to finish a training task and two creation tasks by using VisCommentator.
For the first creation task, 
we curated a raw video \textbf{T1} that contained more than five turns, in which a player lost the rally due to an unforced error, 
and required the experts to augment the video using commentaries in play-by-play style.
The second creation task required the experts to use analyst-style comments on a video \textbf{T2},
in which a player lost the rally due to a forced error in less than five turns.
The training task was prepared to cover all the features in the two creation tasks
by providing a video \textbf{T0} with more than five turns and a player lost due to a forced error.
In total, we prepared six raw videos, each three for tennis and table tennis, respectively.
The original commentaries of each video were removed.
We also provided the experts with a document of the vocabulary and example sentences supported by the system.

\para{Procedure}:
The study began with the introduction (10 min) of 
the study purpose,
the concept of augmented sports videos,
the motivation of language-driven authoring, 
and the concepts in our three-step approach.
Next, we proceeded to the training task (15 min). 
We demonstrated the features of the system with video \textbf{T0} and asked the experts to reproduce the augmented videos themselves.

After the training, we provided the experts with two raw videos (\textbf{T1} and \textbf{T2})
for the two creation tasks (15min for each).
We encouraged the experts to watch and ask questions about the videos before the creation.
For each creation task, we asked the experts to comment on at least three segments of the video 
and to create eight visual effects.
Finally, the session ended with a semi-structured interview (15 min) and a post-study questionnaire (5-Point Likert Scale).
Each session was run in-lab using a 27-inch monitor, following a think-aloud protocol.

\begin{figure}[t]
  \centering
    \vspace{-2mm}
  \includegraphics[width=0.95\columnwidth]{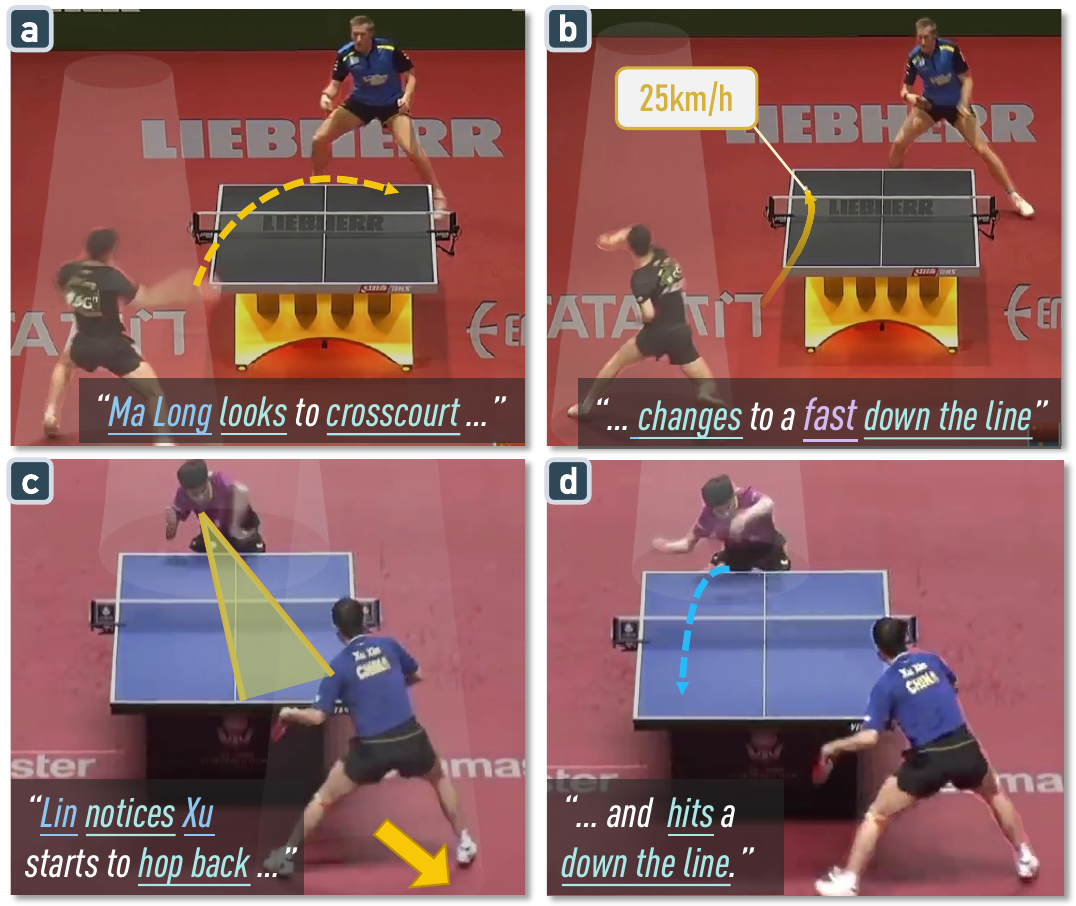}
  \caption{Two examples, a-b and c-d, created by the experts in the study.
  }
 \label{fig:user_example}
 \vspace{-4mm}
\end{figure}

\subsubsection{Results}

\noindent
All experts successfully created multiple augmented videos by using VisCommentator in the creation tasks.
Figure~\ref{fig:user_example} shows two examples of augmented table tennis videos created by the experts during the study.
Unsurprisingly, all experts spoke highly of the usability of the system, considering it \quot{easy to learn and use}.
These results qualitatively demonstrated the efficiency and usability of our system.
The experts’ feedback is summarized as below:

\para{Usefulness}: 
All experts confirmed the usefulness ($\mu = 4.88$) of our language-driven authoring method for sports analysts to create augmented sports videos,
as it \quot{significantly lowers the entry barriers to augmented sports videos for analysts} (P4).
The experts expressed that the usefulness of our method is rooted in its \quot{intuitiveness} and \quot{efficiency} (P1-8),
which allows analysts to create augmented sports videos in a short period without being tangled in the details of video editing. 
Particularly, the experts lauded
that our language-driven method \quot{occupies a unique niche} (P2) of fast prototyping augmented videos 
for day-to-day usages, such as discussion, presentation, and demonstration.
The experts pointed out that in these scenarios,
augmented videos can significantly facilitate the communication of sports insights
but are unnecessary to be high-fidelity.
Thus, existing video editing tools are too ``heavy'' while our method \quot{can perfectly fill this gap} (P8).

\para{Effectiveness}:
The design of our three-step approach was rated as effective by the experts ($\mu = 4.65$).
The experts thought that the four categories of entities purposed by us were \quot{reasonable} (P3-8) 
and \quot{sufficient to present sports insights} (P3, 6-8).
Some experts (P1, 2, 5) suggested that the system should detect some deep semantic meanings, 
such as the players' emotions and the situation of the games.
The experts also agreed with our proposed visualization methods for each category 
and suggested that 
the players' emotions can be visualized by highlighting their actions (P7).
As for the two scheduling methods,
while most of the experts appreciated our proposed designs,
some experts who focused on table tennis (P7-8) thought that play-by-play mode is not that useful
as in table tennis usually both the players and ball move too fast to be augmented.

\para{Satisfaction}:
The rating also reflects positive user satisfaction for the implementation of \system{} ($\mu = 4.23$).
The experts indicated that the detector could correctly recognize the key information in their comments
and the scheduler could properly arrange the visualizations by incorporating both the text orders and video events.
However, comments also suggested that the dictionary that maps entities to
visualizations was not comprehensive enough
so that \quot{I have to find alternative expressions to describe a tactic.} (P6)
We considered this can be mitigated in future improvement by extending the dictionary.

\subsubsection{Observations, Feedback, and Future Opportunities}

\para{Visualizing Deep Semantic Information in Physical Contexts.}
Some experts suggested that the system should be able to detect and visualize
deep semantic meanings in the text.
For example, P6 wanted to visualize the \quot{tense situation} in the game;
P1 noted that although he \quot{comments the same [kinds of] actions}, 
the visualizations should be different since he may have \quot{different tendencies.}
However,
extracting and visualizing the deep semantic information are both challenging tasks,
which requires further study in NLP and visualizations.
From the perspective of visualization,
visualizing such kind of highly abstract information in a physical context remains underexplored.
Recently, research in Situated Visualization~\cite{DBLP:journals/tvcg/BressaKTHV22}, an emerging research topic,
has conducted preliminary exploration in this direction.

\para{Collaborative Interactions Across Abstraction Levels.}
While natural language can allow users to convey high-level insights,
we notice that sometimes the expert wants to express low-level information that can hardly be expressed in language.
For example,
P1 found that it was difficult to express a specific court location in language.
Instead, he would like to type \quot{Federer moves to ...} and then use the mouse to directly click the location in the video.
This interesting observation implies that when authoring augmented videos,
users need interactions with different expressiveness 
ranging from low to high abstraction levels.
However, how to unify the interactions across various abstraction levels remains an open question.
Recently,
Srinivasan et al.~\cite{DBLP:conf/chi/SrinivasanLRDH20} explored consistent multimodal interactions for data visualization on tablets,
providing relevant knowledge in this interesting direction.

\para{Opportunities Enabled by NLIs to Bridge Data Analysis and Communication.}
Surprisingly, the experts suggested that the NLI system not only facilitates the creation of augmented videos but also their analysis process.
P5 provided that the augmentations generated based on comments can be seen as visual notes of the video, 
which helps him externalize and organize his thoughts.
\quot{At the beginning [of analysis,] my thoughts are fragmented..}, P5 detailed that, \quot{...visualizing them can help me to think.}
Such feedback suggests a promising opportunity to bridge the gap between data analysis and communication.
Specifically, with our technique,
a visual analytics system for sports videos 
can enable users to take textual notes on the videos,
visualize their notes to facilitate the analysis,
and gradually shift to authoring augmented sports videos to present the analytical findings.
We discussed this new workflow with the experts
and received very positive feedback.
Thus, we suggest further exploration in this direction.

\para{Suggestions.}
The experts also identified some limitations, most of which were related to system engineering maturity.
For example, 
compositions of multiple augmentations were not supported.
The experts did come across certain issues rooted in the design of our language-driven method.
First, we noticed that the experts were hesitant to type when guessing which words the system could understand.
Such an issue is a long-standing challenge for users of NLI systems~\cite{Discoveringnlp}.
One plausible solution is to integrate auto-complete features into the system.
Second, the experts showed that 
the visual mappings should adapt to the specific sports.
For example, 
the visualizations of distances between a player and a place 
should be different in tennis and table tennis (\emph{i.e.}, stand on the court \emph{vs.} behind the table).
Finally, 
the reading times of entities and the video events
can be mismatched.
The future system should provide interactive functions 
for manual corrections.

\section{Discussion}

\para{Failure Cases Due to ML Models.}
During our study, we observed some failure cases in which \system{} cannot create augmented videos correctly.
One major source of these cases is the imperfect NLP models, especially the SRL model.
For some complex and long sentences, 
such as \quot{Fan, now with the side of the racket that makes it spring off the rubber fast},
the SRL model cannot extract the desired predict-arguments structures,
leading to problematic visualizations.
Moreover, the SRL model can only extract the shallow semantic meanings of the sentence
but cannot understand the deep semantic relationships among entities or sentences.
For example, 
\quot{Dema takes a swatting backhand after this inside-out forehand from Zhendong},
will lead to an error order of what commentators want to convey.
In addition, the underlying CV models may also cause inappropriate rendering results
due to, for example, incorrect object detection, tracking, or segmentation.
Nevertheless,
these issues can be addressed or mitigated with more advanced models, larger datasets, and better implementations~\cite{wu2021ai4vis}.

\para{Generalizability---beyond racket-based sports.}
While \system{} is designed and implemented for tennis and table tennis,
it can be generalized to other racket- or team-based sports 
as it is built based on a formative study of both racket- and team-based sports.
\cmo{Among the three components,
the bottleneck for generalization lies in the Entity Visualizer,
as the other two--Entity Detector (data-driven) and Visualization Scheduler (domain-agnostic)--are naturally generalizable.
Entity Visualizer needs to be extended with a domain specific dictionary that maps sports actions to embedded visualizations.}
Such a verb-visual dictionary will increase the expressive power of \system{},
but also contribute to the visualization community in many directions, such as data animations and NLI systems.
Another bottleneck of extending \system{} to team-based sports is the underlying CV models, 
which require detecting and tracking multiple objects. 
Recent advances in transformer-based models~\cite{gberta_2021_ICML} for video processing can be a solution.

\para{Applicability---broader application scenarios.}
We have showcased that \system{} can be employed in various application scenarios. 
In the user study, the experts also suggested multiple interesting applications.
On one hand, the experts believed that our technique could benefit presenters
in scenarios such as teaching, group discussion, and TV broadcasting.
The experts commented that when presenting insights about sports videos, the generated visualizations 
can \quot{reduce the ambiguity [of spoken language] and facilitate the communication} (P8).
On the other hand, some experts also indicated that our technique can be used in game-viewing systems for audiences.
P8 explained that \quot{fans can type comments to highlight players or visualize data in game watching.}
We consider leveraging interactive embedded visualizations to improve game watching experiences as an interesting future direction.

\para{Study Limitations.}
Since \system{} was built based on findings derived from English spoken commentaries
and was only implemented for tennis and table tennis 
and covered the words discovered from the collected commentaries.
Further studies or adaptations may still be necessary when generalizing it to other scenarios.
Besides, the expert evaluations only provided qualitative feedback
since the sample size is small due to the limited nature of access to experts.
Finally, although the experts were
satisfied with the created augmented videos, 
we didn’t evaluate the videos from the audience’s perspective. 
Follow-up empirical evaluation in real-world settings is thus suggested.
\section{conclusion}
This work aims to facilitate the creation of augmented sports videos using insights expressed in natural language.
To achieve this goal,
we proposed a three-step approach inspired by existing research in text-to-visuals.
We conducted a formative study to analyze \numClip{} augmented sports videos and their commentaries
to answer three key questions in the approach.
Informed by the analysis results,
we designed and implemented \system{},
a proof-of-concept system 
that creates augmented videos for racket-based sports using textual comments.
To demonstrate the applicability of \system{},
we presented two application scenarios, \ie,
authoring augmented sports videos 
and augmenting historical sports videos based on auditory comments.
A technical evaluation showed that \system{} can successfully detect visualizable text entities.
A user study with eight sports analysts revealed the utility, effectiveness, and high satisfaction of the system.
Feedback and observations from the study suggest promising future research directions.

\footnotesize
\acknowledgments{
The authors wish to thank the sports experts from Zhejiang University for their time and expertise.
A special thanks to Salma Abdel Magid for her beautiful voice and help on the video narration.
This research is
supported in part by the 
NSF award III-2107328, NSF award IIS-1901030,
NIH award R01HD104969,
and the Harvard Physical Sciences
and Engineering Accelerator Award.
}

\bibliographystyle{abbrv-doi}

\bibliography{template}
\end{document}